\documentclass{aa}  

\usepackage{graphicx}
\usepackage[hyperindex, breaklinks, colorlinks, linkcolor=blue, citecolor=blue]{hyperref}
\usepackage{txfonts}
\usepackage{calrsfs}
\usepackage{upgreek}
\usepackage{xcolor}
\DeclareMathAlphabet{\pazocal}{OMS}{zplm}{m}{n}

\usepackage{hyperref}

\begin{document}

   \title{Ambipolar diffusion and the mass-to-flux ratio in a turbulent collapsing cloud}

   \author{Aris Tritsis
          \inst{1}}

   \institute{Institute of Physics, Laboratory of Astrophysics, Ecole Polytechnique F\'ed\'erale de Lausanne (EPFL), \\ Observatoire de Sauverny, 1290, Versoix, Switzerland \\
              \email{aris.tritsis@epfl.ch}}
   \date{Received date; accepted date}
   \titlerunning{Ambipolar diffusion and the mass-to-flux ratio in turbulent clouds}
   \authorrunning{A. Tritsis}
 
  \abstract
   {The formation of stars is governed by the intricate interplay of nonideal magnetohydrodynamic (MHD) effects, gravity, and turbulence. Computational challenges have hindered a comprehensive 3D exploration of this interplay, posing a longstanding challenge in our theoretical understanding of molecular clouds and cores.}
   {Our objective was to study both the spatial features and the time evolution of the neutral-ion drift velocity and the mass-to-flux ratio in a 3D chemo-dynamical simulation of a supercritical turbulent collapsing molecular cloud.}
   {Using our modified version of the \textsc{FLASH} astrophysical code, we performed a 3D nonideal MHD simulation of a turbulent collapsing molecular cloud. The resistivities of the cloud were computed self-consistently from a vast non-equilibrium chemical network containing 115 species. To compute the resistivities we used different mean collisional rates for each charged species in our network. We additionally developed a new generalized method for measuring the true mass-to-flux ratio in 3D simulations.}
   {Despite the cloud's turbulent nature, at early times, the neutral-ion drift velocity follows the expected structure from 2D axisymmetric non-ideal MHD simulations with an hourglass magnetic field. At later times, however, the neutral-ion drift velocity becomes increasingly complex, with many vectors pointing outward from the cloud's center. Specifically, we find that the drift velocity above and below the cloud's ``midplane'' is in ``antiphase''. We explain these features on the basis of magnetic helical loops and the correlation of the drift velocity with the magnetic tension force per unit volume. Despite the complex structure of the neutral-ion drift velocity, we demonstrate that, when averaged over a region, the true mass-to-flux ratio monotonically increases as a function of time and decreases as a function of the radius from the center of the cloud. In contrast, the ``observed'' mass-to-flux ratio shows poor correlation with both the true mass-to-flux ratio and the density structure of the cloud.}
   {}


   \maketitle


\section{Introduction}\label{intro}

For the better part of half a century, the role of ambipolar diffusion in the fragmentation of molecular clouds has been one of the most hotly debated topics in the field of star formation (see e.g., \citealt{2006ApJ...646.1043M, 2010MNRAS.402L..64C, 2010MNRAS.409..801M, 2012ARA&A..50...29C, 2015MNRAS.451.4384T, 2019MNRAS.488.2357P, 2019FrASS...6....7K, 2022FrASS...9.9223M} and referenes therein). This lack of consensus can be partially attributed to substantial challenges that render the exploration of such effects, both theoretically and observationally, exceedingly difficult. 

Thus far, only tentative indications of the neutral-ion drift velocity have been found in observations \citep{2008ApJ...677.1151L, 2010ApJ...720..603H, 2018ApJ...862...42T}. These studies concluded that neutral species have a higher velocity dispersion than charged ones. On the other hand, \cite{2021ApJ...912....7P, 2024A&A...690L...5P} reached the opposite conclusion, and reported that the charged species exhibited a higher velocity dispersion than the neutrals. Additionally, they found no evidence that the power spectrum of ions has more power at smaller scales. Three key factors hinder our ability to observationally study ambipolar diffusion. Firstly, the neutral-ion drift velocity typically found in numerical simulations \citep{2001ApJ...550..314D, 2022MNRAS.510.4420T} is a factor of two to three smaller than the typical spectral resolution of most spectroscopic surveys. Secondly, the observed charged and neutral species used to study ambipolar diffusion need to be co-spatial in order for the difference in the spectral lines to be attributed to the neutral-ion drift velocity \citep{2012ApJ...760...57T}. Finally, the optical depth of the two species needs to be comparable so that the derived neutral-ion drift velocity is not biased by opacity-broadening and other radiative-transfer effects. \cite{2023MNRAS.521.5087T} extensively studied all these effects and, by employing 2D axisymmetric nonideal magnetohydrodynamic (MHD) chemo-dynamical simulations and non-local thermodynamic equilibrium (non-LTE) radiative-transfer simulations, presented a roadmap for future observations of ambipolar diffusion.

From the theoretical perspective, it is certain that nonideal MHD effects have to set in at some stage during the collapse of molecular clouds. This is essential to explain the considerably weaker magnetic field observed in young stars compared to the magnetic field that would emerge from a prestellar core collapsing under flux-freezing (e.g.,\citealt{2015MNRAS.452..278T, 2018A&A...615A...5V}). Additionally, the onset of nonideal MHD effects is inevitable in the densest parts of molecular clouds where the degree of ionization is of the order of $\chi_i = 10^{-6}-10^{-9}$ \citep{1998ApJ...503..689W, 1999ApJ...512..724B, 2002ApJ...565..344C, 2009A&A...498..771G}.

To date, several 3D simulations have incorporated nonideal MHD effects, often however, employing varying levels of simplifications in their treatment of these processes \citep{2015MNRAS.452..278T, 2016A&A...587A..32M, 2016A&A...592A..18M, 2017A&A...603A.105D, 2018A&A...615A...5V, 2018MNRAS.475.1859W, 2020A&A...635A..67H, 2021MNRAS.505.5142Z, 2024ApJ...970...41M, 2024A&A...682A..30L}. Most of these calculations focus on the role of nonideal MHD effects in protostellar disks, while only a few are targeted in prestellar cores (e.g., \citealt{2021MNRAS.504.2381Y}). Additionally, other simulations which focus on earlier stages during the star formation process are usually limited in terms of dimensionality \citep{2009NewA...14..483B, 2010MNRAS.408..322K, 2012ApJ...753...29T, 2012ApJ...760...57T, 2022MNRAS.510.4420T, 2023MNRAS.521.5087T}

Incorporating nonideal MHD effects into our calculations is critical to study variations (both in time and space) of the mass-to-magnetic flux ratio, one of the key parameters in star-formation theories \citep{1976ApJ...210..326M}. Observationally, many studies have attempted to explore such variations\footnote{By ``time variations'' in the context of observations, we refer to comparisons of measured mass-to-flux ratios across different clouds and cores, which are assumed to be at different evolutionary stages.} (e.g., \citealt{2008A&A...487..247F, 2009ApJ...692..844C, 2012ApJ...747...80K, 2012ARA&A..50...29C, 2021ApJ...913...85H, 2021ApJ...912..159P, 2021ApJ...917...35M, 2022Natur.601...49C, 2023ApJ...942...32Y}). While all these studies have provided valuable insights, only those utilizing Zeeman observations can measure the relevant component of the magnetic field, needed to calculate the flux. The magnetic flux is defined as the dot product between the magnetic field and the normal vector to a given surface. Polarization observations trace the plane-of-sky (POS) component of the magnetic field, while the surface element $\textit{\textbf{ds}}$ is along the line-of-sight (LOS). Since the two vectors are orthogonal, the magnetic flux measured by polarization observations is, by definition, zero. In other words, there is no flux of the POS component of the magnetic field through the POS. We note that this is not merely a mathematical ``formality'' but a fundamental constraint that cannot be ignored. However, even for Zeeman observations, the question remains as to how well they can provide reliable information of the mass-to-flux ratio given 3D effects (e.g., \citealt{2009MNRAS.400L..15M}).

Here, we present results from a nonideal MHD simulation of a collapsing turbulent molecular cloud where the resistivities are self-consistently calculated from all charge carriers, the abundances of which are computed by a vast non-equilibrium chemical network. More than four million CPU hours and $\sim$2.5 calendar years were required for this simulation to reach a maximum number density of $10^5~\rm{cm^{-3}}$. We used this simulation to study the structure of the neutral-ion drift velocity and that of the mass-to-flux ratio. 

Our paper is organized as follows: In Sect.~\ref{numer} we outline the numerical methods we used to perform the nonideal MHD simulation and the physical setup of the collapsing cloud. In Sect.~\ref{SpatialAD} we present first results from our simulation on the neutral-ion drift velocity for two different times during the evolution of the cloud. In Sect.~\ref{MFMain} we discuss our results on the true ``3D'' mass-to-flux and demonstrate that the ``observed'' mass-to-flux ratio that would be inferred from Zeeman observations shows a poor correlation with the true values. Finally, in Sect.~\ref{discuss}, we provide a summary of our findings and conclude.

\section{Numerical methods and setup}\label{numer}

To perform our simulations we largely follow the numerical methods outlined in \cite{2022MNRAS.510.4420T, 2023MNRAS.521.5087T, 2025A&A...695A..18T, 2025A&A...696A..35T} with a few key improvements both in relation to the calculation of the resistivities and in relation to the chemical modeling.

\subsection{nonideal MHD}\label{equations}

Under nonideal MHD conditions, the magnetic induction equation is
\begin{equation}\label{induc}
\frac{\partial\boldsymbol{B}}{\partial t} = \boldsymbol{\nabla}\times(\boldsymbol{v_n}\times\boldsymbol{B}) - c\boldsymbol{\nabla}\times(\eta_\perp \boldsymbol{j_\perp} + \eta_\parallel \boldsymbol{j_\parallel} + \eta_H	\boldsymbol{j}\times \boldsymbol{b}).
\end{equation}
where $\boldsymbol{B}$, $\boldsymbol{b}$, and $\boldsymbol{v_n}$ denote the magnetic field, the unit vector of the magnetic field and the velocity of the neutrals, respectively. In Eq.~\ref{induc}, $\boldsymbol{j}$ is the current, and $\boldsymbol{j_\parallel}$ and $\boldsymbol{j_\perp}$ are the components of the current parallel and perpendicular to the magnetic field. In turn, $\eta_\perp$, $\eta_\parallel$, and $\eta_H$ are the perpendicular, parallel, and Hall resistivities, defined in terms of the respective conductivities, $\sigma_\perp$, $\sigma_\parallel$, and $\sigma_H$ \citep{1991pspa.book.....P}, as 
\begin{gather} 
\qquad \eta_\parallel = \frac{1}{\sigma_\parallel},
\qquad
\eta_\perp = \frac{\sigma_\perp}{\sigma_\perp^2 + \sigma_H^2},
\qquad
\eta_H = \frac{\sigma_H}{\sigma_\perp^2 + \sigma_H^2}.
\label{etas}
\end{gather} 
The conductivities are defined as
\begin{equation}\label{sigmas}
\begin{gathered} 
\qquad \qquad \sigma_\parallel = \sum_s\sigma_s, 
\qquad
\sigma_\perp = \sum_s\frac{\sigma_s}{1+(\omega_s\uptau_{sn})^2}, \\
\qquad \sigma_H = -\sum_s\frac{\sigma_s\omega_s\uptau_{sn}}{1+(\omega_s\uptau_{sn})^2},
\end{gathered}
\end{equation}
where $\omega_s$ and $\sigma_s$ are, respectively, the gyrofrequency and conductivity of species ``$s$'', defined as
\begin{equation}\label{gyrofreq}
\omega_s = \frac{e_sB}{m_sc},
\end{equation}
\begin{equation}\label{sigs}
\sigma_s = \frac{n_se_s^2\uptau_{sn}}{m_s},
\end{equation}
where $n_s$ and $m_s$ are the number density and mass of charged species $s$, $c$ is the speed of light, and $e_s$ is the charge of each species. We note that we only considered singly-ionized species. However, the symbol $e_s$ carries the algebraic sign of the charge of each species considered. In Eqs.~\ref{sigmas} and \ref{sigs}, $\uptau_{sn}$ is the mean collisional time between species $s$ and neutral particles, defined as \citep{1996ASIC..481..505M}
\begin{equation}\label{mct}
\uptau_{sn} = \frac{1}{\alpha_{s\rm{He}}}\frac{m_s+m_{\rm{H_2}}}{\rho_{\rm{H_2}}}\frac{1}{\overline{\sigma w}_{s\rm{H_2}}},
\end{equation}
where $m_{\rm{H_2}}$ and $\rho_{\rm{H_2}}$ are the mass and mass density of $\rm{H_2}$, $\overline{\sigma w}_{s\rm{H_2}}$ is the mean collisional rate of species $s$ with $\rm{H_2}$, and $\alpha_{s\rm{He}}$ is an additional factor to account for the deceleration of species $s$ resulting from collisions with $\rm{He}$. For ions the mean collisional rate can be obtained from the Langevin approximation. In contrast to \cite{2022MNRAS.510.4420T} and \cite{2023MNRAS.521.5087T}, we did not use constant values for $\overline{\sigma w}_{s\rm{H_2}}$ and $\alpha_{s\rm{He}}$ for all ions. Instead, the mean collision rate was individually determined for each ion as \citep{1961ApJ...134..270O, 2008A&A...484...17P}
\begin{equation}\label{mcr}
\overline{\sigma w}_{s\rm{H_2}} = 2.21\pi \Big(\frac{e_s^2p_{\rm{H_2}}}{\mu}\Big)^{1/2},
\end{equation}
where $\mu$ is the reduced mass of species $s$ and $\rm{H_2}$, and $p_{\rm{H_2}}$ is the polarizability of $\rm{H_2}$. For the factor $\alpha_{s\rm{He}}$ we adopted \citep{2008A&A...484...17P}
\begin{equation}\label{alphaf}
\alpha_{s\rm{He}} = 1 + \Bigg[\frac{(m_s+m_{\rm{H_2}})m_{\rm{He}} p_{\rm{He}}}{(m_s+m_{\rm{He}})m_{\rm{H_2}} p_{\rm{H_2}}}\Bigg]^{1/2} \Bigg(\frac{n_{\rm{He}}}{n_{\rm{H_2}}}\Bigg),
\end{equation}
where $p_{\rm{He}}$ and $m_{\rm{He}}$ are the polarizability and mass of $\rm{He}$, and $n_{\rm{H_2}}$ and $n_{\rm{He}}$ are the number densities of $\rm{H_2
}$ and $\rm{He}$. Here, we adopted $p_{\rm{H_2}} = 8.04 \times 10^{-25}~\rm{cm^{3}}$ and $p_{\rm{He}} = 2.07 \times 10^{-25}~\rm{cm^{3}}$ \citep{1961ApJ...134..270O}. We note that the maximum variation in collisional rate coefficients among different ion species can reach up to $\sim$67\% (see, e.g., Figs. 5 and 7 from \citealt{2008A&A...484...17P}). In turn, this variation (along with variations in $\alpha_{s\rm{He}}$) translates into differences in the perpendicular resistivity of $\lesssim$15\%\footnote{The relative error in the perpendicular resistivity is not constant (both spatially and as the cloud evolves) but instead depends on the relative contribution of each species to the perpendicular conductivity which changes in time and space (e.g., Fig 6 from \citealt{2022MNRAS.510.4420T}).} compared to, say, adopting a constant value for all species. While this is not a drastic deviation at any given time, it is a systematic one that can accumulate over the course of tens of thousands of integration steps. Our initial elemental abundances are listed in Table 1 of \cite{2022MNRAS.510.4420T}. For electrons, the Langevin approximation is not applicable. \cite{1996ASIC..481..505M} extrapolated the results by \cite{1971ToAC} to temperatures relevant to molecular clouds, and found a value for the mean collisional rate between electrons and $\rm{H_2}$ of $\overline{\sigma w}_{eH_2}$ = 1.3$\times$$10^{-9}$ $\rm{cm^{3}}$~$\rm{s^{-1}}$. For the factor $\alpha_{eHe}$, we adopt a value of 1.16. Finally, for grains we have that $\alpha_{g^-He}$ = 1.28 \citep{2005ApJ...618..769T} and $\overline{\sigma w}_{g^-H_2}$ is given by \citep{1924PhRv...23..710E, 1993ApJ...418..774C, 2008A&A...484...17P}
\begin{equation}\label{mcrgrains}
\overline{\sigma w}_{g^-H_2} = \frac{4}{3}\pi r_g^2 \delta(8k_B\mathrm{T}/\pi m_{H_2})^{1/2},
\end{equation}
where $k_B$ is the Boltzmann constant, T is the temperature, $r_g$ is the grain radius and $\delta$ is a dimensionless parameter determined from laboratory experiments to be equal to 1.3 \citep{2003PhPl...10....9L}. We adopted an MRN grain-size distribution \cite{1977ApJ...217..425M} following the methods outlined in \cite{2009MNRAS.399L..94K} and considering thirty bins of grain radii with $r_g^{min}$ = 0.0181 $\rm{\upmu m}$ and $r_g^{max}$ = 0.9049 $\rm{\upmu m}$. Additionally, we adopted a constant grain abundance of $n_{g^-}/n_{\rm{H_2}} = 10^{-12}$ \citep{2012ApJ...753...29T} and a mass density of $\rm{\rho_{g^-}}$ = 2.3 g $\rm{cm^{-3}}$. A growing body of theoretical and observational studies suggest that grain growth could begin prior to the protostellar phase (e.g., \citealt{2009A&A...502..845O, 2010A&A...511A...9S, 2010Sci...329.1622P, 2014A&A...567A..32M, 2019A&A...632A...5G, 2019MNRAS.488.4897V, 2020A&A...643A..17G, 2022MNRAS.514.2145B}). At the same time, during the early phases of molecular cloud evolution, grains can contribute up to $\sim$10\% to the perpendicular conductivity, depending on the underlying assumptions adopted. While considering grain coagulation and/or a different grain size distribution could moderately influence our results, adopting an MRN grain-size distribution remains a reasonable assumption for the range of densities considered here.

We note here that the Langevin approximation is valid as long as the neutral-ion drift velocity ($\rm{\boldsymbol{v_{dr}}}$) is less than the sound speed \cite{1996ASIC..481..505M}. However, if magnetic-field gradients become steep enough, this assumption may be invalidated \citep{2024arXiv240506026H}. In our implementation we accounted for such discrepancies using the following approach. For every grid point we computed the resistivities and drift velocities based on the Langevin rates for the mean collisional rates. If the drift-velocity in the grid point was supersonic, we computed new mean collisional rates taking into account the dependence of the mean collisional rate on the drift velocity itself (see e.g., Figs. 3, 5, and 20 from \citealt{2008A&A...484...17P}). To calculate the new collisional rates for the ions and the grains we used the relations provided by \cite{2008A&A...484...17P, 2008A&A...492....1P} whereas for electrons we use Eq. (14) from \citeauthor{1983ApJ...264..485D} (\citeyear{1983ApJ...264..485D}; see also \citealt{1979ems..book...81P}), adjusted such that, when the electron-neutral drift velocity is zero, the value of the mean collisional rate was $\overline{\sigma w}_{eH_2}$ = 1.3$\times$$10^{-9}$ $\rm{cm^{3}}$~$\rm{s^{-1}}$, as stated above. With these new collisional rates, we calculated new values for the resistivities. Finally, the process continued iteratively until the resistivity values converged to within a 1\% tolerance. Given the weak dependence of the mean collisional rates on the drift velocities and the fact that the drift velocities in our simulation only become marginally supersonic, only very minor adjustments on the final values of the resistivities were observed.

\subsection{Chemical network}

The inclusion of a chemical network is imperative to accurately calculate the resistivities. Specifically, the chemical network was used to compute the number densities of species $s$ (i.e., $n_s$ that first appeared in Eq.~\ref{sigs}) in every point in space and time. We use a non-equilibrium gas-grain chemical network to model the evolution of 115 species described in \citeauthor{2025A&A...695A..18T} (\citeyear{2025A&A...695A..18T, 2025A&A...696A..35T}; see also, \citealt{2016MNRAS.458..789T, 2022MNRAS.510.4420T, 2023MNRAS.521.5087T}). We adopt reaction rates from the fifth release of the \textsc{UMIST} database \citep{2013A&A...550A..36M}. The network consisted of $\sim$1650 chemical reactions and included a ranges of processes from freeze-out to cosmic-ray and photo-related processes. The visual extinction that is used to determine the reaction rates of photo-related processes was calculated ``on the fly'' for every grid point based on the six-ray approximation, following a similar approach as \cite{2012MNRAS.421..116G}. For the cosmic-ray ionization rate we adopted the ``standard'' value of $\zeta_0 = 1.3\times10^{-17}~\rm{s^{-1}}$ \citep{1998ApJ...499..234C}. Given its role in setting the ionization fraction, the cosmic-ray ionization rate is a critical free parameter in non-ideal MHD simulations. Even modest variations, as small as a factor of two, can significantly impact the dynamical evolution of a cloud, as well as the magnitude and spatial structure of the drift velocity (see e.g., Figs. 1 and 3 from \citealt{2023MNRAS.521.5087T}). Further observational efforts, such as those by \citet{2023ApJ...947L..18S}, are therefore crucial for placing tighter constraints on its value.

\subsection{Numerical setup}

To perform our simulation of a collapsing turbulent molecular cloud, we made use of the adaptive mesh refinement (AMR) code \textsc{FLASH} \citep{2000ApJS..131..273F, 2008ASPC..385..145D}. We used an identical numerical and physical setup as in \cite{2025A&A...695A..18T, 2025A&A...696A..35T} with the only exception being the inclusion of nonideal MHD effects and a different initial condition for the value of the magnetic field. Below, we provide only a brief description of our setup.

Our initial values for the number density and magnetic-field strength were set equal to 500 $\rm{cm^{-3}}$ and $10~\rm{\mu G}$, respectively. Given our initial strength for the magnetic field, which was initially oriented along the $z$ axis, the value of mass-to-flux ratio (in units of the critical value for collapse \citealt{1976ApJ...210..326M}), was $\sim$1.7. We assumed isothermality, with the temperature of the cloud being set equal to $\rm{T} = 10~K$. The dimensions of our simulated cloud were equal to 2 pc, we used open boundary conditions, and the initial size of our grid was $64^3$ grid points. However, we also included two levels of AMR refinement and, as a result, the resolution of our smallest cell was $\sim8\times10^{-3}$ pc. The total mass in our simulation box was 240 $\rm{M_\odot}$. Turbulence was also initiated as described in \cite{2025A&A...695A..18T}, based on the publicly-available code \texttt{TurbGen} (\citealt{2022ascl.soft04001F}; see also \citealt{2010A&A...512A..81F}). Given our initial values for the temperature and strength of the magnetic field, the initial sonic and Alfv\'en Mach numbers were $\mathcal{M_\mathrm{s}}=\sigma_v/c_\mathrm{s}\sim3$ and $\mathcal{M_\mathrm{A}}=\sigma_v/v_\mathrm{A}\sim0.85$, respectively. We emphasize that by adopting the initial conditions described above, we do not imply that these necessarily represent the typical initial conditions of real molecular clouds. Instead, this simulation is part of a broader suite designed to systematically explore the impact of different parameters on the dynamics and chemistry.

\begin{figure*}
\hspace*{0.5cm}
\includegraphics[width=2.\columnwidth, clip]{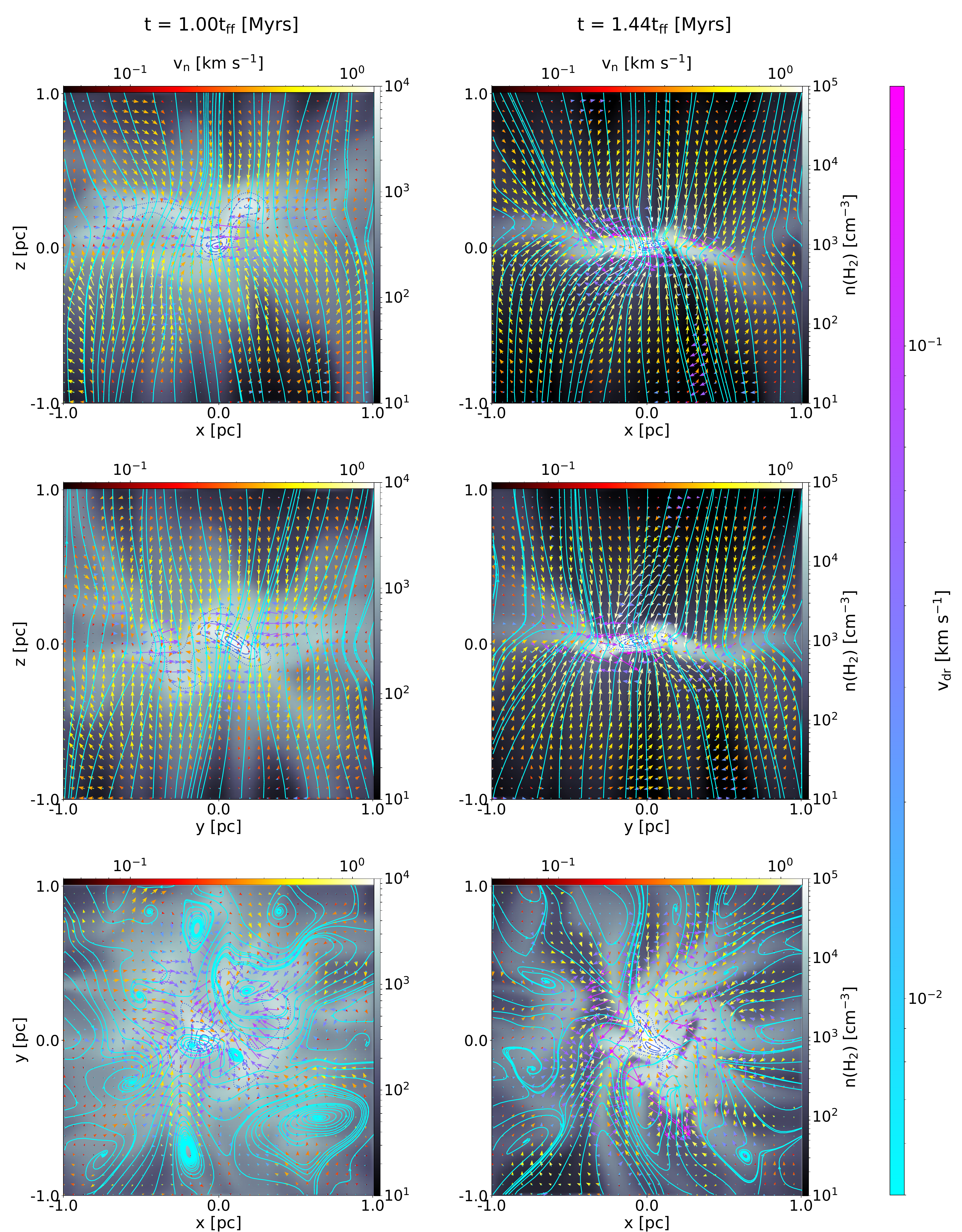}
\caption{Density slices (black-white colormap) through our computational cube at the location of the maximum density at one free-fall time (left column) and at the end of our simulation (right column). From top to bottom we show slices through the $y$, $x$ and $z$ directions. Cyan streamlines show the morphology of the magnetic field in each slice and blue contours mark the density with dotted, dashed, dash-dotted and solid lines marking, respectively, the 30, 50, 70, and 90 \% of the maximum density. The red-yellow colormap in each panel and respective arrows show the velocity of the neutrals in the simulation and the cyan-fuchsia colormap and respective arrows show the neutral-ion drift velocity.
\label{Slices}}
\end{figure*}

\section{Results}\label{Results}

\subsection{The spatial structure and time evolution of the neutral-ion drift velocity}\label{SpatialAD}

In Fig.~\ref{Slices} we present $\rm{H_2}$ number density slices through the computational cube perpendicular to the $y$ (upper row), $x$ (middle row), and $z$ (bottom row) directions, respectively. In the left column we present our results for one free-fall time (hereinafter $t_{ff}$), and in the right column we show our results at the end of our simulation, corresponding to 1.44 $t_{ff}$. Slices are taken at the location of the maximum density. The overplotted cyan lines show the morphology of the magnetic field for each slice and the blue contours mark the 30, 50, 70, and 90 \% of the maximum density in this specific slice (dotted, dashed, dash-dotted, and solid lines, respectively). The velocity field of the neutrals is shown with the red-yellow arrows and respective colormaps, and the neutral-ion drift velocity is shown with the cyan-fuchsia arrows.

From Fig.~\ref{Slices} several notable features are visible. Firstly, the neutral-ion drift velocity peaks in the surroundings of the densest regions of the cloud rather than precisely at the location of the maximum density. This trend stands for both times during the evolution of the cloud examined here. For purely gravitationally-driven ambipolar diffusion, the neutral-ion drift velocity should scale as $v_{dr}\propto -\uptau_{ni\perp} \boldsymbol{g_\perp}$, where $\boldsymbol{g_\perp}$ is the component of the gravitational field perpendicular to magnetic field lines \citep{1987ASIC..210..491M}. Given that $\boldsymbol{g_\perp}$ vanishes at the center of the cloud, so should the neutral-ion drift velocity. Additionally, to first order, the perpendicular resistivity scales as $\eta_\perp\propto v_a^2/n_{i, tot}$, where $v_a$ is the Alfv\'en velocity and $n_{i, tot}$ is the total number density of all ions (see Eq. 11 from \citealt{2025A&A...695A..82A}). At a time of 1.44 $t_{ff}$, both the total ion density and the Alfv\'en velocity decrease by a similar factor ($\sim$5-7) as the density increases from few times $10^4~\rm{cm^{-3}}$ to $10^5~\rm{cm^{-3}}$. However, since the perpendicular resistivity depends on the square of the Alfv\'en velocity, the reduction in $v_a$ has a stronger impact on $\eta_\perp$ than the change in total ion density. Additionally, the ``midplane'' (e.g., $z\approx0$) of the cloud, where the density reaches its maximum value, is a zero-crossing point for the perpendicular components of the current. Since $v_{dr} \propto \eta_\perp J_\perp$ (see Eq.~\ref{driftVelEqs}), the drift velocity is suppressed in the densest region and peaks in the surroundings. Finally, the low-density regions of the cloud ($\sim 10^2-10^3~\rm{cm^{-3}}$) remain close to the ideal MHD regime, as expected.

The neutral-ion drift velocity is also found to be primarily perpendicular to the velocity of the neutrals. This can be understood as follows. Along magnetic field lines the neutral-ion drift velocity is nearly zero since both the ions and the neutrals are subject to the same forces. Therefore, the only significant velocity separation occurs for $\rm{\boldsymbol{v_{dr, \perp}}}$. Given that, for the most part, the velocity of the neutrals follows magnetic-field lines, the drift velocity is perpendicular to both the velocity of the neutrals and the magnetic field. This is particularly evident in regions where the magnetic field has the most pronounced hourglass morphology.

From the left column of Fig.~\ref{Slices} it can also be seen that the neutral-ion drift velocity is in very good agreement with the expectations for the gravitationally-driven ambipolar diffusion from 2D axisymmetric simulations. That is, the vectors of the neutral-ion drift velocity point inward, toward the center of the cloud with this being the case both above and below the midplane of the cloud. This is also evident in the $z$ slice through our computational cube (lower left panel in Fig.~\ref{Slices}) where, again, the majority of the vectors of the neutral-ion drift velocity point toward the densest regions of our simulated cloud. However, even at this stage during the evolution of the cloud, a few vectors are more randomly oriented compared to the case of a simple isolated collapsing core with an hourglass morphology. Consequently, some kinetic (or ``gravo-kinetic'') ambipolar diffusion is also present, although this does not seem to be the dominant effect at this stage. A radial profile of the magnitude of the drift velocity perpendicular to the mean direction of the magnetic field (not shown here) also looks remarkably similar to 2D axisymmetric nonideal MHD simulations.

Despite this very good agreement with gravitationally-driven ambipolar diffusion at early stages during the evolution of the cloud, the situation is drastically different at later times. Especially in the slice perpendicular to the $z$ axis and for $t = 1.44~t_{ff}$ (bottom right panel in Fig.~\ref{Slices}), the neutral-ion drift velocity exhibits a very complicated structure. Of particular interest are the two elongated structures at $x\approx-0.2$ pc (with $y\approx-1 - 0$ pc), and $x\approx-0.1$ pc (with $y\approx 0.2- 0.75$ pc) created by solenoidal motions where the magnetic field also exhibits a highly pinched morphology. In both these regions, the neutral-ion drift velocity is oriented perpendicular to the long axis of these elongated structures, suggesting that the neutrals are dispersing outward from the spine of these filamentary structures compared to the ions. However, the most striking feature from the right column of Fig.~\ref{Slices} is the ``messy'' structure of the neutral-ion drift velocity in the $x$ and $y$ slices. That is, many vectors point outward instead of inward, toward the densest parts of the cloud. Additionally, the direction of the neutral-ion drift velocity above and below the miplane of the cloud, is in ``antiphase''. 

\begin{figure}
\includegraphics[width=1.\columnwidth, clip]{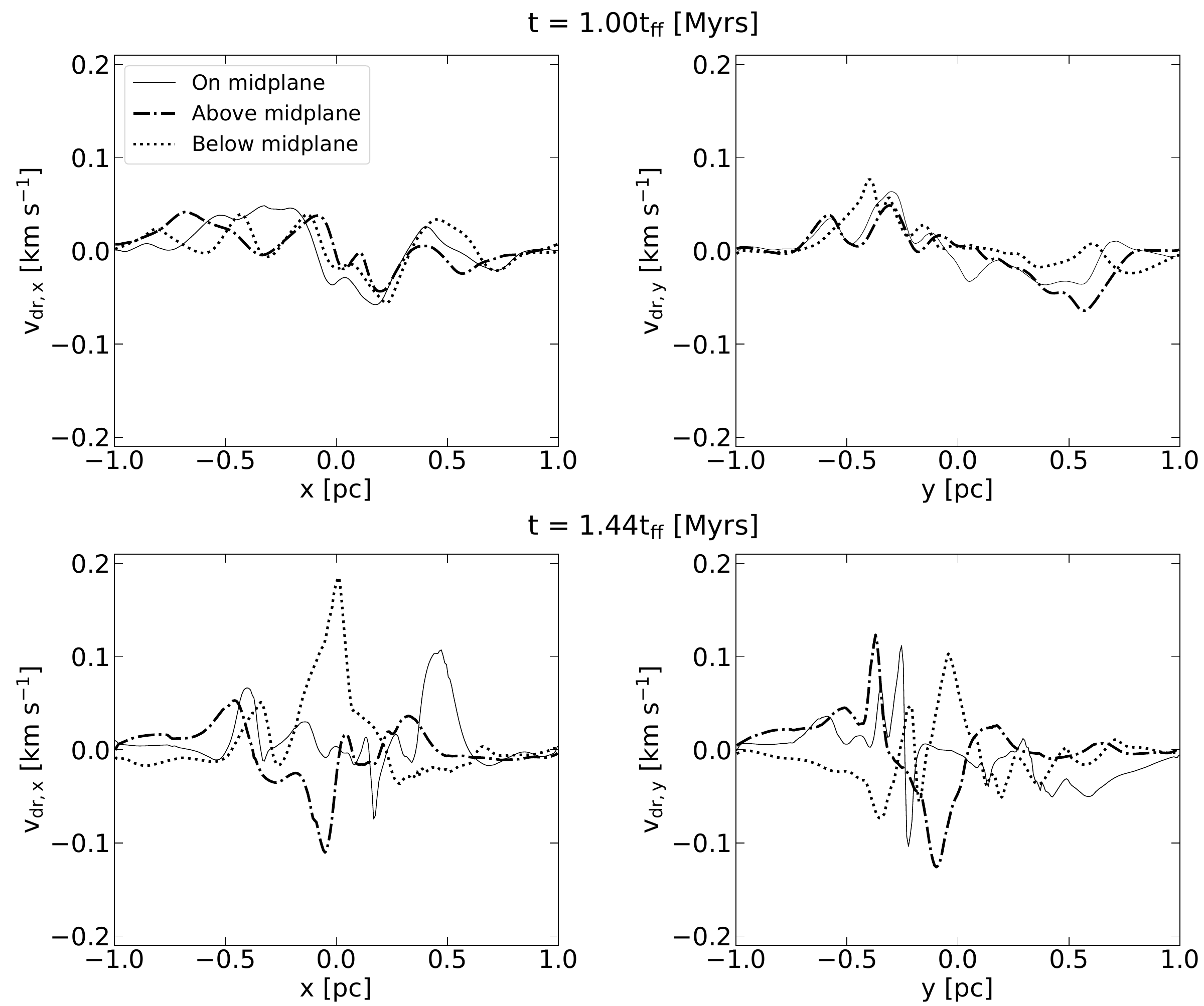}
\caption{One-dimensional profiles of the $x$ and $y$ components of the neutral-ion drift (left and right panels, respectively) for two different times during the evolution of the cloud. In the upper row, we present our results for one free-fall time and in the bottom row we show our results at the end of the simulation. With the thin solid lines, we show the 1D profiles on the midplane of the cloud (i.e., $z \approx 0$), and with the dash dotted and dotted lines we show the profiles above and below the midplane. Notice the approximate anticorrelation of the profiles above and below the midplane at the end of simulation.
\label{VdrProfs}}
\end{figure}

To further emphasize this point, we show in Fig.~\ref{VdrProfs} 1D profiles of the $x$ and $y$ components of the neutral-ion drift velocity (left and right columns, respectively) for $t = 1.~t_{ff}$ and $t = 1.44~t_{ff}$ (upper and lower rows, respectively). With the thin solid lines we show profiles on the midplane of the cloud and with the thick dash-dotted and dotted lines we show profiles above and below the midplane, respectively. All 1D profiles were obtained by averaging over a 0.05 pc slice in the $z$ direction. The profiles above and below the midplane were taken at a vertical distance of 0.1 pc from the midplane. As is evident from the bottom row of Fig.~\ref{VdrProfs}, at a time of $t = 1.44~t_{ff}$, the 1D profiles of the drift velocity above and below the midplane are approximately in antiphase. We have confirmed, that this is the case in other cuts through our computational cube.

\begin{figure}
\includegraphics[width=1.\columnwidth, clip]{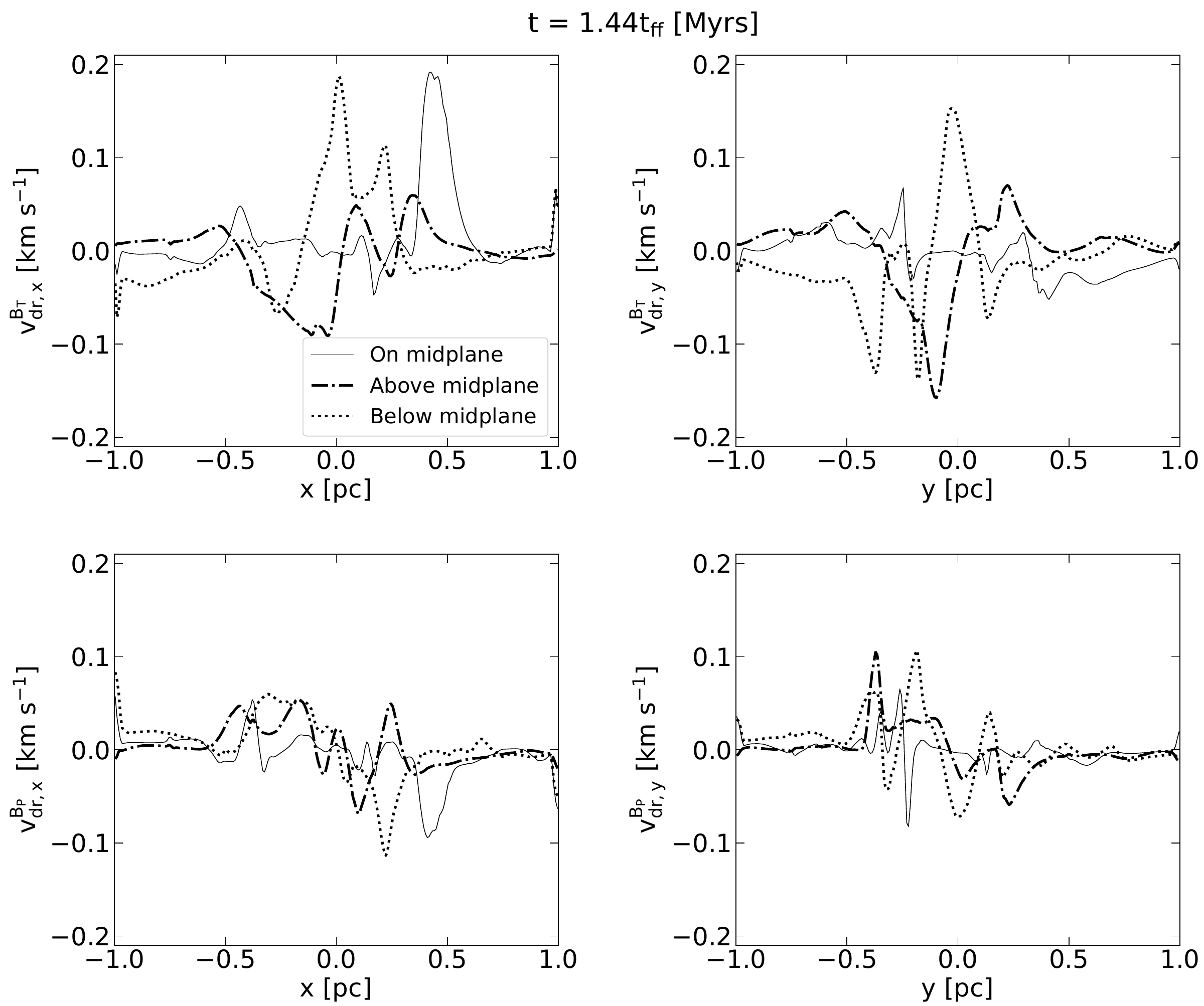}
\caption{One-dimensional profiles of the $x$ and $y$ components of the neutral-ion drift (left and right panels, respectively) at the end of the simulation. In the upper row, we show the components of the neutral-ion drift velocity driven by the magnetic tension force per unit volume, and in the lower row we show the components driven by the magnetic pressure force per unit volume. Note that, to zeroth order, the features observed in the lower row of Fig.~\ref{VdrProfs} can be explained by the component of the neutral-ion drift velocity driven by the magnetic tension force per unit volume.
\label{VdrProfsDecomp}}
\end{figure}

To identify the physical mechanism behind this approximate anticorrelation, we decompose the drift velocity into two components: one driven by the magnetic tension force per unit volume and one driven by the magnetic pressure force per unit volume (see Appendix~\ref{vdrEqs}). We note that this separation does not imply that one component is exclusively associated to gravitationally driven ambipolar diffusion and the other with kinetic ambipolar diffusion, as both gravity and turbulence can bend and/or compress the magnetic field. In Fig.~\ref{VdrProfsDecomp}, we show these two components of the drift velocity for a time of $1.44~t_{ff}$. In the upper row we show the $x$ and $y$ components of the drift velocity (left and right panels, respectively) driven by the magnetic tension force per unit volume. In the lower row we show the corresponding components of neutral-ion drift velocity associated to the magnetic pressure force per unit volume. The linestyles have the same meaning as in Fig.~\ref{VdrProfs}. From Fig.~\ref{VdrProfsDecomp} we can draw two conclusions. Firstly, the components of the neutral-ion drift velocity associated with the magnetic tension force per unit volume are notably higher (i.e., by a fact of $\gtrsim$2). Secondly, the approximate anticorrelation of the drift velocity above and below the midplane of the cloud is primarily driven by the magnetic tension force per unit volume.

\begin{figure}
\includegraphics[width=1.\columnwidth, clip]{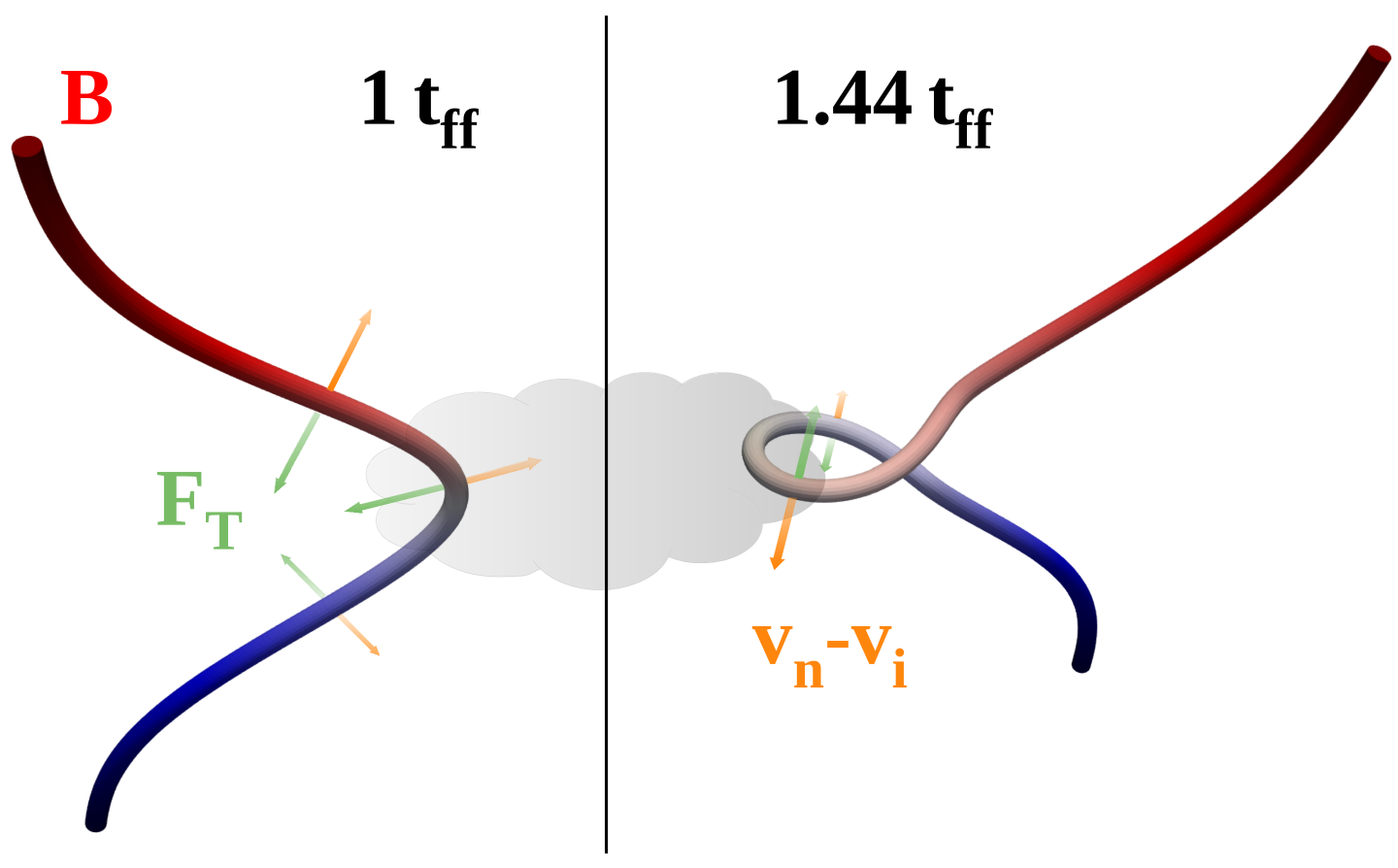}
\caption{Illustration of the physical picture that explains (to zeroth order) the features seen in the drift velocity in Fig.~\ref{Slices} at two different times during the evolution of the cloud. Color-coded tubes represent magnetic field lines, green arrows show the direction of the magnetic tension force per unit volume, and orange arrows show the direction of the neutral-ion drift velocity. On the left side of the plot we depict the situation at one free-fall time, when magnetic-field lines have approximately an hourglass morphology. The right side of the plot shows the evolved magnetic-field morphology (at 1.44 $t_{ff}$), which explains the anti-correlation seen in the neutral-ion drift velocity with respect to the midplane of the cloud. 
\label{Schematic}}
\end{figure}

Upon inspecting the 3D structure of the magnetic field in our simulation at a time of $1.44~t_{ff}$, we identified that a significant portion of magnetic field lines exhibit helical ``loops'' near the midplane of the cloud. These helical loops are in addition to the hourglass morphology which is clearly evident in the upper right, and middle right panels of Fig.~\ref{Slices}. In turn, such helical structures arise in regions of high vorticity (see also the velocity of the neutrals in the bottom right panel of Fig.~\ref{Slices}). In such magnetic field morphologies, the vector of the magnetic tension force per unit volume points toward the inner regions of the loop, both above and below the midplane. If we ignore any ``contributions'' by the magnetic pressure force per unit volume to the neutral-ion drift velocity, the latter must have an opposite sign to the magnetic tension force per unit volume. Therefore, the neutral-ion drift velocity should point in one direction above the midplane and in the opposite direction below the midplane. This leads to the anticorrelation of the 1D profiles observed in Fig.~\ref{Slices}--\ref{VdrProfsDecomp}. Additionally, such helical loops lead to a situation where a significant portion of vectors of the neutral-ion drift velocity do not point toward the inner regions of the cloud.

This physical picture is schematically explained on the right side of Fig.~\ref{Schematic}. With the green arrows we show the direction of the magnetic tension force per unit volume and with the orange vectors we show the direction of the neutral-ion drift velocity. The morphology of the magnetic field lines is depicted with the color-coded tubes. To avoid cluttering, we have not indicated the direction of the magnetic tension force at the region of the helical loop closer to the center of the cloud. The magnetic tension force in this location would point away from the center of the cloud, leading to a drift velocity pointing toward the cloud's center. Hence, some vectors of the neutral-ion drift velocity should still point inward. On the left side of Fig.~\ref{Schematic}, we show the structure of the neutral-ion drift velocity (and that of the magnetic tension force per unit volume) expected from a simple hourglass morphology in the magnetic field (consistent with 1 $t_{ff}$). We emphasize here again, that we have ignored any contributions from the magnetic pressure force per unit volume which are, in reality, non-negligible, and can further complicate the picture.

\subsection{The mass-to-flux ratio}\label{MFMain}

\subsubsection{Average properties of the mass-to-flux ratio in time and space}

The neutral-ion drift velocity is of paramount importance in the evolution of the cloud, as it governs the spatial variations of the mass-to-flux ratio. Specifically, the mass-to-flux ratio should increase in the same direction as the direction of the neutral-ion drift velocity. The particularly messy morphology of the neutral-ion drift velocity at later times, and the fact that a significant fraction of the vectors point outward from the dense region raise an important and provocative question: ``could the chaotic structure of the neutral-ion drift velocity lead to a reduction of the mass-to-flux ratio in the densest region?''

To answer this question, we developed a new method for measuring the true 3D mass-to-flux ratio. By true mass-to-flux ratio, we refer to what is also known in the literature as the differential mass-to-flux ratio (e.g., \citealt{1991ApJ...373..169M}). Our approach is described in Appendix~\ref{trueMF}. In Fig.~\ref{AveragedMFTime} we show the time evolution of the mass-to-flux ratio in a region centered on the location of the maximum density at a time of 1.44 $t_{ff}$, and spanning 0.2 pc in the $x$ and $y$ directions. We then follow the evolution of the mass-to-flux ratio in this region through time. The black points represent the 50th percentile from all pixels in the region and the errorbars represent the 16th and 84th percentiles, respectively. With the red horizontal line we depict our initial condition for the mass-to-flux ratio. As is evident from Fig.~\ref{AveragedMFTime}, the mass-to-flux ratio in the region monotonically increases, with only a very tentative indication for a slight reduction of the median value from all pixels at very late times. To examine to what extent this slight reduction depends on our choice of the region and averaging effects, we repeat the procedure outlined above, this time selecting a zoomed-in region that spans 0.1 pc in the $x$ and $y$ directions, and it is again centered on the location of the maximum density at a time of 1.44 $t_{ff}$. We present our results with the red points in Fig.~\ref{AveragedMFTime}. In this zoomed-in region the monotonic increase of the mass-to-flux ratio as a function of time is even more clear.

We now proceed to examine the spatial features of the mass-to-flux ratio at a time of 1.44 $t_{ff}$. To this end, we select a larger region spanning, 0.6 pc in the $x$ and $y$ directions and is again centered on the location of the maximum density. We then create radial profiles of the true mass-to-flux ratio and the density by considering concentric radial shells. We present our results on the true mass-to-flux ratio with the blue points in the upper panel of Fig.~\ref{AveragedMF}. The points represent the 50th percentile from all pixels in the shell and the errorbars represent the 16th and 84th percentiles, respectively. Our results on the radial profile of the density are shown in the bottom panel of Fig.~\ref{AveragedMF} (radial profiles of the mass-to-flux ratio and the density at a time of 1 $t_{ff}$ are presented in Appendix~\ref{RadProfs1tff}). As is evident, the average mass-to-flux ratio is in good agreement with the 1D density profile in that it peaks in the same location as the density and gradually declining outward, until it reaches a value close to the initial condition. Therefore, based on Figs.~\ref{AveragedMFTime} and \ref{AveragedMF} we conclude that, at least in terms of its average properties, the complex structure of the neutral-ion drift velocity cannot lead to a reduction in the mass-to-flux ratio.

Despite this conclusion, we note two important points. Firstly, there seem to be additional small features in the radial profile of the mass-to-flux ratio (i.e., at $r\approx0.1$ pc) which are not present in the radial profile of the density. Additionally, other features (i.e., at $r\approx0.2$ pc) seem to have an offset with respect to the density profile. However, given our systematic uncertainties on the individual pixel level (see the discussion in Appendix~\ref{bench}), it remains uncertain whether these minor discrepancies hold any significance. Secondly, and most importantly, while the average 1D profiles of the mass-to-flux ratio and density align well, their peak locations in the full 2D maps do not coincide. We will further explore this issue in a future study.

To understand why the mass-to-flux ratio does not decrease at late stages, we quantify the direction of the drift velocity in the cloud. First, we consider slices perpendicular to the $z$ axis in the range $\mid z \mid \le 0.25$ pc. In each slice, we define a reference point ($x_m, ~y_n$), corresponding to the location of maximum density in the midplane slice. For every pixel within a given slice, we construct a vector pointing toward ($x_m, ~y_n$) and compute the angle $\theta_\measuredangle$ between this vector and the local drift velocity in the $x-y$ plane. For simplicity, we set the $z$ component of the drift velocity to zero, as vertical motions will be ultimately regulated by gravity. In other words, the important physical process here is in which direction the neutrals will cross magnetic field lines in different planes perpendicular to the mean component of the field.

We present our results on $\theta_\measuredangle$ in Fig.~\ref{AnglePDFs}. With the black and red lines we show the probability density functions of the $\theta_\measuredangle$ for a time of 1 $t_{ff}$ and 1.44 $t_{ff}$, respectively. $\theta_\measuredangle = 0^\circ$ signifies that the drift velocity points toward the dense region, and $\theta_\measuredangle = 180^\circ$ means that the drift velocity points away from the cloud's center. As is evident from Fig.~\ref{AnglePDFs}, at 1 $t_{ff}$ the vast majority of the vectors of the drift velocity point towards the center of the cloud, as expected from the simple physical picture shown on the left side of Fig.~\ref{Schematic}. At later times (red line in Fig.~\ref{AnglePDFs}), a significant fraction of vectors point outward from the cloud's center. Yet, the majority of the vectors still point inward or at $\theta_\measuredangle \approx 90^\circ$. This can again be explained based on the the simple physical picture shown on the right side of Fig.~\ref{Schematic}. We note that $\theta_\measuredangle \approx 90^\circ$ should not significantly affect the mass-to-flux ratio. A combination of drift velocities oriented at $\theta_\measuredangle \approx 90^\circ$ across different locations would result in a ``flow'' pattern resembling a circular vortex causing neutrals to ``circulate'' around the central region (relative to the ions), rather than escape it. Our results do not qualitatively change if we consider fewer slices and/or if we consider only the drift velocities with magnitude above some threshold (e.g., $\|v_{dr}\| \ge 0.01~\rm{km~s^{-1}}$).  

\begin{figure}
\includegraphics[width=1.\columnwidth, clip]{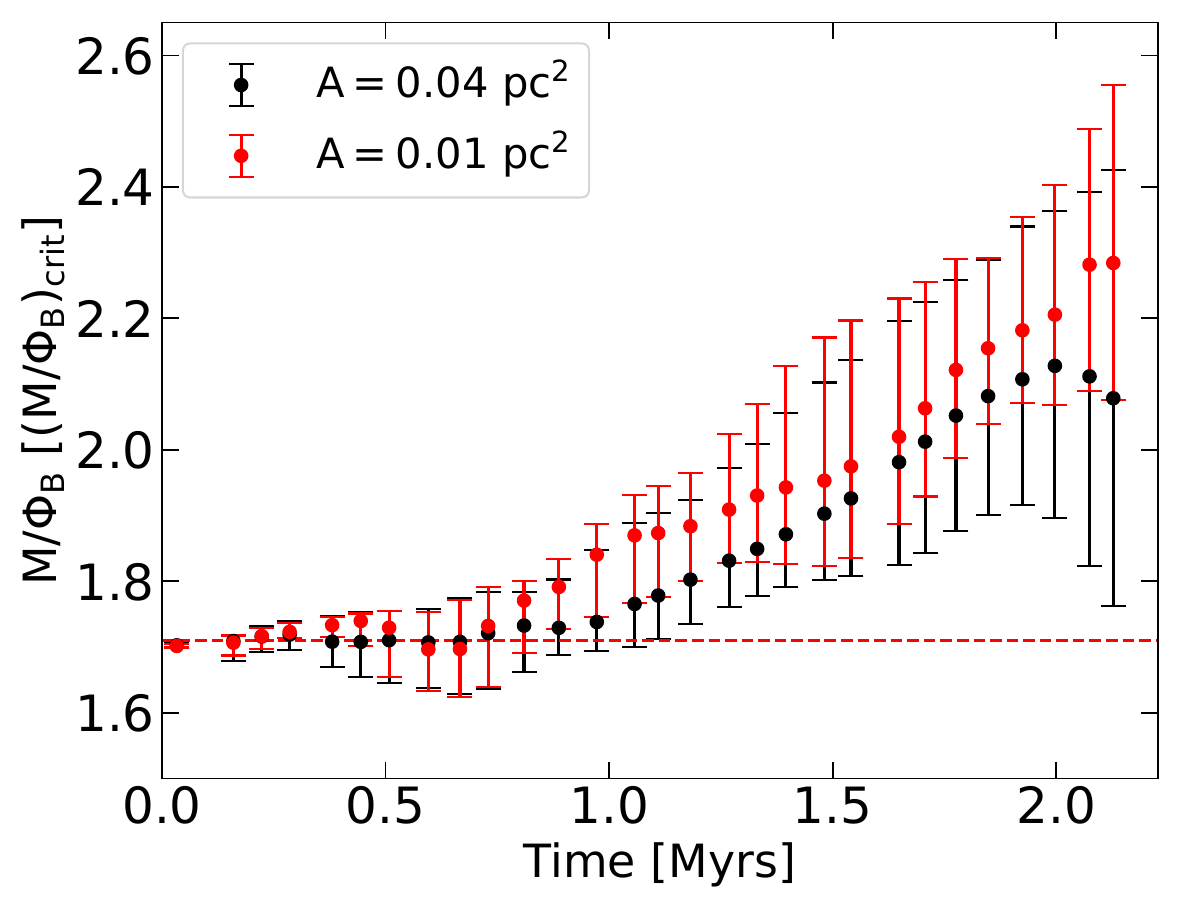}
\caption{The evolution of the mass-to-flux ratio as a function of time in a slice perpendicular to the $z$ axis, centered on the location of the maximum density at a time of 1.44 $t_{ff}$. The mass-to-flux ratio was computed for all cells within a 0.2$\times$0.2 $\rm{pc^2}$ area. With the black points we show the 50th percentile from all pixels in the region and the errorbars represent the 16th and 84th percentiles, respectively. With the red points we have zoomed further in to the location where the density eventually increases. The area of this region is 0.01 $\rm{pc^2}$. The red dashed line marks the initial value of the mass-to-flux ratio in the simulation.
\label{AveragedMFTime}}
\end{figure}

\subsubsection{The ``observed'' mass-to-flux ratio}\label{obsMF}

Examining all the complications that enter in observational measurements of the mass-to-flux ratio through Zeeman observations requires non-LTE spectropolarimetric radiative transfer calculations of species such as $\rm{CN}$ and/or $\rm{OH}$. Even though both these species are present in our chemical network, we avoid such complications and consider the following idealized scenario. Firstly, we assume that the cloud is viewed along the mean direction of the magnetic field (i.e., along the $z$ axis of our simulation). Secondly, we assume that our ``Zeeman observations'' can perfectly probe the LOS component of the magnetic field (i.e. the $z$ component) at the midplane of the cloud, where the density of the cloud peaks. With these assumptions we essentially consider a hypothetical molecule that is perfectly concentrated at the midplane of the cloud, and we ignore any complications arising from chemical and/or radiative transfer effects. We then measure the observed mass-to-flux ratio as described in Appendix~\ref{mockMF}. As such, our results should be considered the best-case scenario regarding how well we can probe the mass-to-flux ratio with Zeeman observations.

\begin{figure}
\includegraphics[width=1.\columnwidth, clip]{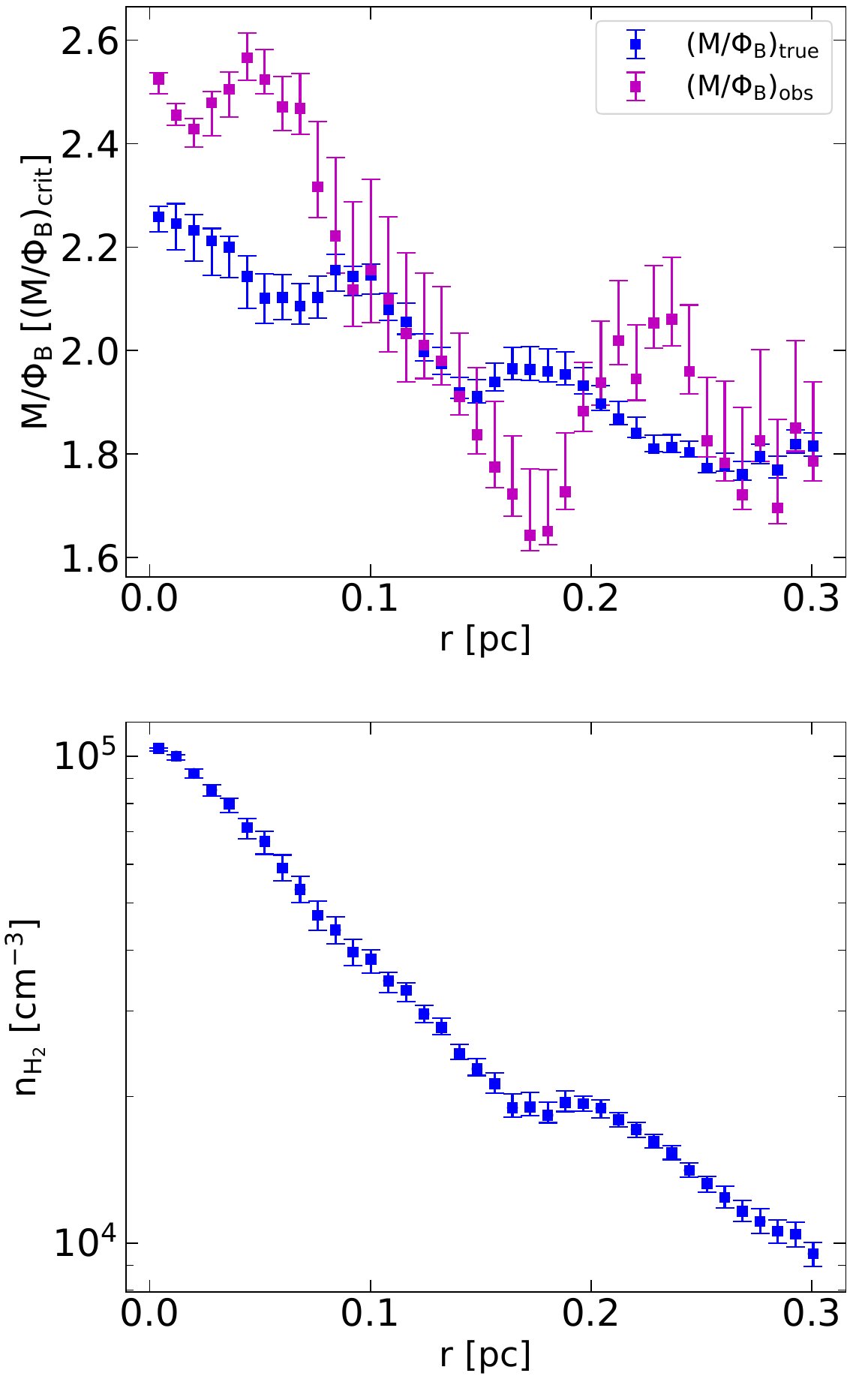}
\caption{Average profiles of the mass-to-flux ratio (upper panel), and the number density (lower panel) of the cloud in concentric radial shells at a time of 1.44 $t_{ff}$. The averages were created by considering a slice perpendicular to the $z$ axis centered at the location of the maximum density. With the blue points in the upper panel we show the true mass-to-flux ratio, and with the magenta points we show the observed mass-to-flux ratio. Even though the drift velocity exhibits complex features, there is still good correlation in terms of averages between the true mass-to-flux ratio and the number density of the cloud. On the other hand, the radial profile of the observed mass-to-flux ratio does not correspond well to neither the true mass-to-flux ratio nor to the number density of the cloud.
\label{AveragedMF}}
\end{figure}

\begin{figure}
\includegraphics[width=1.\columnwidth, clip]{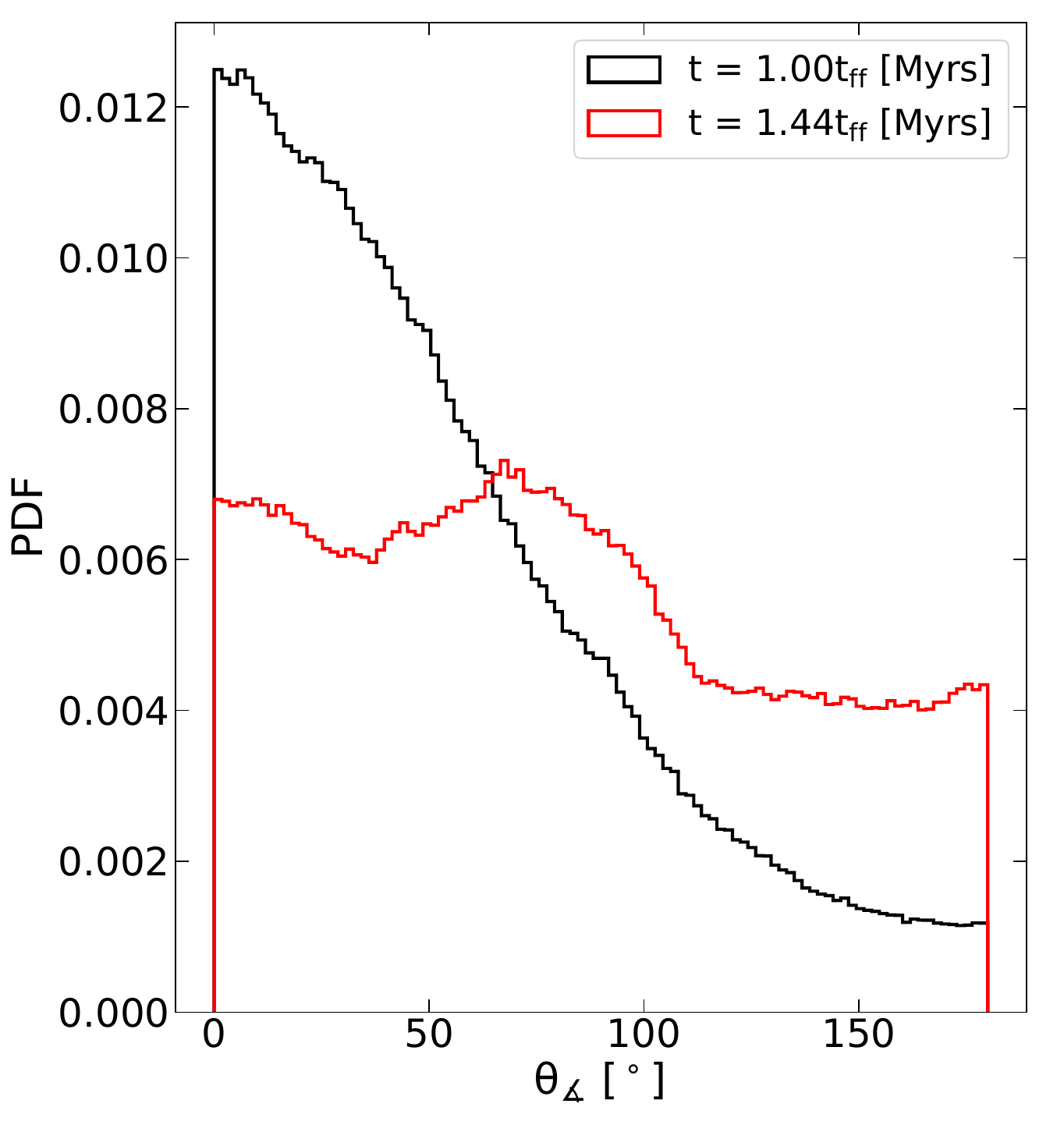}
\caption{Probability density functions of the angle between the drift velocity and vectors pointing toward the center of the cloud for the two different times examined. $\theta_\measuredangle = 0^\circ$ means that the neutral-ion drift points toward the center of the cloud and $\theta_\measuredangle = 180^\circ$ means that the drift velocity points away from the dense region.
\label{AnglePDFs}}
\end{figure}

With the magenta points in the upper panel of Fig.~\ref{AveragedMF}, we show the radial profile of the observed mass-to-flux ratio averaged in concentric radial shells. The points and errorbars have the same meaning as in the case of the true mass-to-flux ratio. As evident, the observed mass-to-flux ratio does not accurately reflect the true values, overestimating them in some regions and underestimating them in others. Most notably, the observed mass-to-flux ratio does not monotonically decrease outward. That could lead into a scenario where, if the beam of a telescope was placed at $r\approx0.24$ pc, and at $r\approx0.17$ pc, the mass-to-flux ratio would appear to be decreasing toward the center of the core \citep{2009ApJ...692..844C}. These discrepancies arise because of the fact that, observationally, we cannot follow the true 3D structure of magnetic flux tubes or, in other words, we cannot perform a line integral along a flux tube. Therefore, this is a geometrical effect, unrelated to nonideal MHD effects and/or turbulence, that needs to be considered when interpreting observations. We have also confirmed that using a thicker slice (i.e., averaging the magnetic field over multiple layers close to the midplane of the cloud, instead of using a single plane) does not qualitatively alter our results and the correlation of the observed mass-to-flux ratio with the true values does not improve. Finally, we note that the same discrepancies between the observed mass-to-flux ratio and the true one can also be realized in H\textsc{I} narrow self-absorption Zeeman observations \citep{2022Natur.601...49C} for the same reason as outlined above.

\section{Summary and conclusions}\label{discuss}

We performed a nonideal MHD chemo-dynamical simulation of a supercritical transAlfv\'enic supersonic collapsing molecular cloud. The resistivities were self-consistently calculated from the abundances of the charge carriers in our chemical network. We studied the structure of the neutral-ion drift velocity in two different times during the evolution of the cloud. We developed a novel approach for measuring the true mass-to-flux ratio in MHD simulations. Using this method, we examined the time evolution of the mass-to-flux ratio, and computed radially averaged profiles from the cloud's center to assess global trends.

At early times, the structure of the neutral-ion drift velocity is in very good agreement to 2D axisymmetric nonideal MHD models that have a simple hourglass magnetic field morphology. That is, the vast majority of the vectors of the neutral-ion drift velocity point towards the center of the cloud, with this being the case both above and below the cloud's midplane. However, as the cloud evolves, the structure of the drift velocity becomes significantly more complicated. Specifically, we found that the drift velocity above and below the cloud's midplane is in anti-phase. Additionally, we analyzed the angle between the drift velocity and the direction toward the center of the cloud. We found that a substantial fraction of the drift velocity vectors point away from the center of the cloud, albeit the number of vectors pointing inward is still significantly higher. Finally, we found that a another significant fraction of neutrals, cross magnetic field lines at angles of $\theta_\measuredangle \approx 90^\circ$ relative to the direction toward the cloud's center. We show that this behavior of the drift velocity at later times is caused by the magnetic tension force per unit volume, and the presence of magnetic helical loops, which are formed in high vorticity regions. Based on these results, we conclude that the presence of turbulence can significantly affect the structure of the neutral-ion drift velocity, further challenging the interpretation of observations aiming to quantify ambipolar diffusion.

Despite these complexities, we demonstrate that, when averaged over a central region, the true mass-to-flux ratio still follows the expected monotonic increase over time, and a radial decrease towards the edges of the cloud. This is substantial evidence that the mass accumulation (relative to the magnetic flux) in the central parts of the cloud is not disrupted by the turbulence-induced changes in the structure of the drift velocity, and that the predictions of the ambipolar diffusion theory of star formation \citep{1999ASIC..540..305M} are still relevant.

Finally, we show that the observed mass-to-flux ratio, even in an idealized scenario and even upon averaging over a region, fails to accurately capture the spatial variations of the true mass-to-flux ratio and shows poor correlation with the underlying density structure of the cloud. This discrepancy highlights the challenges in interpreting observational measurements and the necessity to account for 3D effects in Zeeman measurements of the mass-to-flux ratio.

\begin{acknowledgements}

We thank K. Tassis, T. Mouschovias, S. Basu, C. Federrath, J. Schober, G. V. Panopoulou, R. Skalidis, and V. Andrianatou for stimulating discussions and comments. We also thank the referee for a useful report that helped us improve the manuscript. A. Tritsis acknowledges support by the Ambizione grant no. PZ00P2\_202199 of the Swiss National Science Foundation (SNSF). The software used in this work was in part developed by the DOE NNSA-ASC OASCR Flash Center at the University of Chicago. This research was enabled in part by support provided by SHARCNET (Shared Hierarchical Academic Research Computing Network) and Compute/Calcul Canada and the Digital Research Alliance of Canada. We also acknowledge use of the following software: \textsc{Matplotlib} \citep{2007ComputSciEng.9.3}, \textsc{Numpy} \citep{2020Nat.585..357}, \textsc{Scipy} \citep{2020NatMe..17..261V} and the \textsc{yt} analysis toolkit \citep{2011ApJS..192....9T}.

\end{acknowledgements}

\begin{appendix}

\section{The neutral-ion drift velocity}\label{vdrEqs}

By combining the momentum equation for charged species with the generalized Ohm's law (see \citealt{2023MNRAS.521.5087T} for a detailed derivation), it can be shown that

\begin{eqnarray}\label{driftVelEqs}
\begin{bmatrix}
\frac{m_{\rm{s}}c}{\uptau_{\rm{sn}}e_s} & B_z & -B_y \\
-B_z & \frac{m_{\rm{s}}c}{\uptau_{\rm{sn}}e_s} & B_x \\
B_y & -B_x & \frac{m_{\rm{s}}c}{\uptau_{\rm{sn}}e_s}
\end{bmatrix}
\begin{bmatrix}
v_{{\rm{dr}}, x} \\
v_{{\rm{dr}}, y} \\
v_{{\rm{dr}}, z}
\end{bmatrix} = c
\begin{bmatrix}
\eta_\perp j_{\perp, x} + \eta_\parallel j_{\parallel, x} + \eta_{\rm{H}}(\boldsymbol{j}\times \boldsymbol{b})_x \\
\eta_\perp j_{\perp, y} + \eta_\parallel j_{\parallel, y} + \eta_{\rm{H}}(\boldsymbol{j}\times \boldsymbol{b})_y \\
\eta_\perp j_{\perp, z} + \eta_\parallel j_{\parallel, z} + \eta_{\rm{H}}(\boldsymbol{j}\times \boldsymbol{b})_z
\end{bmatrix},
\end{eqnarray}
where all of the symbols have the same meaning as in Sect.~\ref{equations}. Additionally, by neglecting the terms containing the parallel and Hall resistivities, as well as other subdominant terms it follows that
\begin{equation}\label{vdrifts}
\begin{gathered} 
\qquad v_{dr,x} = -\frac{c\eta_\perp}{B^2}(\boldsymbol{j}\times\boldsymbol{B})_x, 
\qquad
v_{dr,y} = -\frac{c\eta_\perp}{B^2}(\boldsymbol{j}\times\boldsymbol{B})_y, \\
\qquad v_{dr,z} = -\frac{c\eta_\perp}{B^2}(\boldsymbol{j}\times\boldsymbol{B})_z, 
\end{gathered}
\end{equation}
which can, in turn, be written as
\begin{equation}\label{vdrifts2}
\begin{gathered} 
v_{dr,x} = -\frac{c^2\eta_\perp}{4\pi B^2}\Big((\boldsymbol{B}\cdot\nabla)\boldsymbol{B}\Big)_x+ \frac{c^2\eta_\perp}{8\pi B^2}\Big(\nabla B^2\Big)_x, \\
v_{dr,y} = -\frac{c^2\eta_\perp}{4\pi B^2}\Big((\boldsymbol{B}\cdot\nabla)\boldsymbol{B}\Big)_y+ \frac{c^2\eta_\perp}{8\pi B^2}\Big(\nabla B^2\Big)_y, \\
v_{dr,z} = -\frac{c^2\eta_\perp}{4\pi B^2}\Big((\boldsymbol{B}\cdot\nabla)\boldsymbol{B}\Big)_z+ \frac{c^2\eta_\perp}{8\pi B^2}\Big(\nabla B^2\Big)_z.
\end{gathered}
\end{equation}
Therefore, the neutral-ion drift velocity can be decomposed into two components; one associated to the magnetic tension force per unit volume and the other to the magnetic pressure force per unit volume.

\section{Measuring the mass-to-flux ratio}\label{MFProcessSect}

\subsection{Calculating the true mass-to-flux ratio}\label{trueMF}

\begin{figure}
\includegraphics[width=1.\columnwidth, clip]{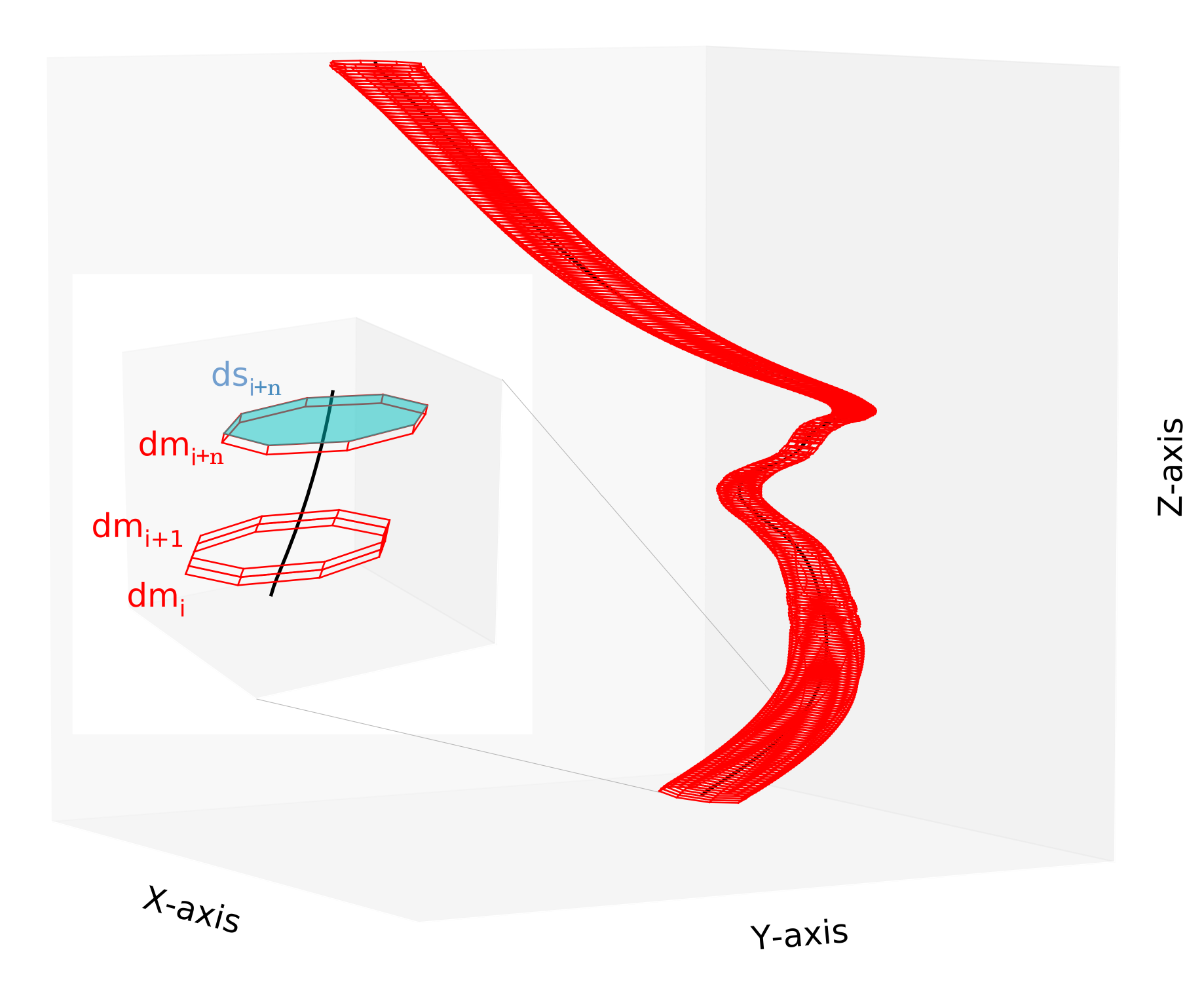}
\caption{Schematic representation of the process followed to measure the mass-to-flux ratio. First, we follow magnetic field lines in our simulation box (black line). Using some basis vectors in each set of positions ``$i$'' and ``$i+1$'' along a magnetic field line, we define convex regions as shown in the inset panel. The volume of these regions is calculated using Delaunay triangulation and the density is interpolated from data in the simulation. The surface ``$ds_i$'', used to compute the magnetic flux, is also calculated using triangulation (see text for more details). 
\label{MFProcess}}
\end{figure}

The basic procedure to calculate the true mass-flux-ratio is schematically described in Fig.~\ref{MFProcess}. We begin by defining a slice perpendicular to the $z$ axis of our simulation at $z = z_k$. Each grid cell ($x_i$, $y_j$) within this slice defines an initial position used to trace magnetic field lines through the computational domain. Integration is performed using a fourth-order accurate Runge-Kutta method \citep{RK4}. Since the magnetic-field-line tracing involves arbitrary positions ($x', y', z'$) that may fall in between the cell centers, we use trilinear interpolation to determine the magnetic field components at these locations while preserving the divergence-free condition\footnote{https://github.com/jpomoell/pysmsh}. The integration process continues until a traced point is beyond the simulation domain. The same procedure is repeated in the opposite direction (i.e., for $z < z_k$) by reversing the field direction.

Once we have the points $r_n$ that trace a magnetic field line, we compute the orthogonal unit vectors $\hat{\mathbf{e}}_{1,i}$, $\hat{\mathbf{e}}_{2,i}$, and $\hat{\mathbf{e}}_{3,i}$, at a position ``$i$'' along the magnetic field line as
\begin{equation}
\hat{\mathbf{e}}_{1,i} = \frac{\mathbf{B}(r_i)}{\|\mathbf{B}(r_i)\|}, \quad
\hat{\mathbf{e}}_{2,i} = \frac{\begin{pmatrix} -e_{1y,i} \\ e_{1x,i} \\ 0 \end{pmatrix}}{\sqrt{e_{1x,i}^2 + e_{1y,i}^2}}, \quad
\hat{\mathbf{e}}_{3,i} = \hat{\mathbf{e}}_{1,i} \times \hat{\mathbf{e}}_{2,i}.
\end{equation}
We then extend outward from the field line in the directions given by the vectors $\pm\hat{\mathbf{e}}_{2,i}$, $\pm\hat{\mathbf{e}}_{3,i}$ and the diagonal directions $\pm(\hat{\mathbf{e}}_{2,i} + \hat{\mathbf{e}}_{3,i})/\sqrt{2}$ and $\pm(\hat{\mathbf{e}}_{2,i} - \hat{\mathbf{e}}_{3,i})/\sqrt{2}$. The same process is repeated for the next location ``$i+1$'' along the magnetic field line. To scale these displacements appropriately, we introduce the scaling factors $f_i$ and $f_{i+1}$, which depend on the local magnetic field strength. These factors ensure that the cross-sectional shape adapts to variations in the field strength along the field line (see Fig.~\ref{MFProcess}), thereby also ensuring that the magnetic flux is constant regardless of the location along the field line. The maximum allowed displacement is one cell.

Through this process we have a set of 16 vertices that define a convex shape with volume $dV_i$. The volume $dV_i$ of these shapes is calculated using Delaunay triangulation to break the region into tetrahedra, with the volume of each tetrahedron being
\begin{equation}
V_{tetrahedron} = \frac{1}{6}|\mathrm{det}(a-d, b-d, c-d)|,
\end{equation}
where $a, b, c$, and $d$ are the vertices of each individual tetrahedron. The mass density inside each convex shape $\rho_{x',y', z'}$ is calculated using trilinear interpolation. Having computed the mass density and the volume of the shape in question, we then simply compute $dm_i$. The total mass loading of the flux tube is calculated by repeating the process for all positions along the magnetic field line and adding all the infinitesimal mass elements, as shown in the inset panel of Fig.~\ref{MFProcess}. The area $ds_i$, used in the calculation of the magnetic flux, is again calculated using triangulation. The magnetic flux is then computed as
\begin{equation}
\Phi_B = \mathbf{B}(r_i)\cdot\hat{\mathbf{B}}(r_i)ds_i,
\end{equation}
and, as mentioned, is constant regardless of the location at which is measured along an individual magnetic field line. 

\subsection{Benchmarking the approach for measuring the mass-to-flux ratio}\label{bench}

The process introduced in Appendix~\ref{trueMF} to measure the true mass-to-flux ratio may be subject to several sources of errors. The most significant of such errors stem from the accuracy with which we can trace magnetic field lines. This becomes increasingly challenging as the cloud is collapsing and the magnetic field morphology gets more complex. Another source of error is related to the accuracy with which we can calculate the volume of the individual volume elements. Interpolation in both the magnetic field and the density, can also introduce small errors in the calculation of the individual mass elements, which can however accumulate, especially in regions of the cloud with strong density and/or magnetic-field gradients. Finally, exporting the simulation data into a uniform grid also introduces some errors.

To verify and benchmark our approach for calculating the mass-to-flux ratio, and access the significance of these errors, we analyze the ideal MHD simulation described in \cite{2025A&A...695A..18T, 2025A&A...696A..35T}. In contrast to the simulation described in the present paper, that simulation was supercritical (with a mass-to-flux ratio of 2.28), but was performed using the same base resolution and the same levels of refinement. Given that no nonideal MHD effects were introduced in the simulation performed by \cite{2025A&A...695A..18T, 2025A&A...696A..35T} we expect to recover the initial mass-to-flux ratio of 2.28.

\begin{figure}
\includegraphics[width=1.\columnwidth, clip]{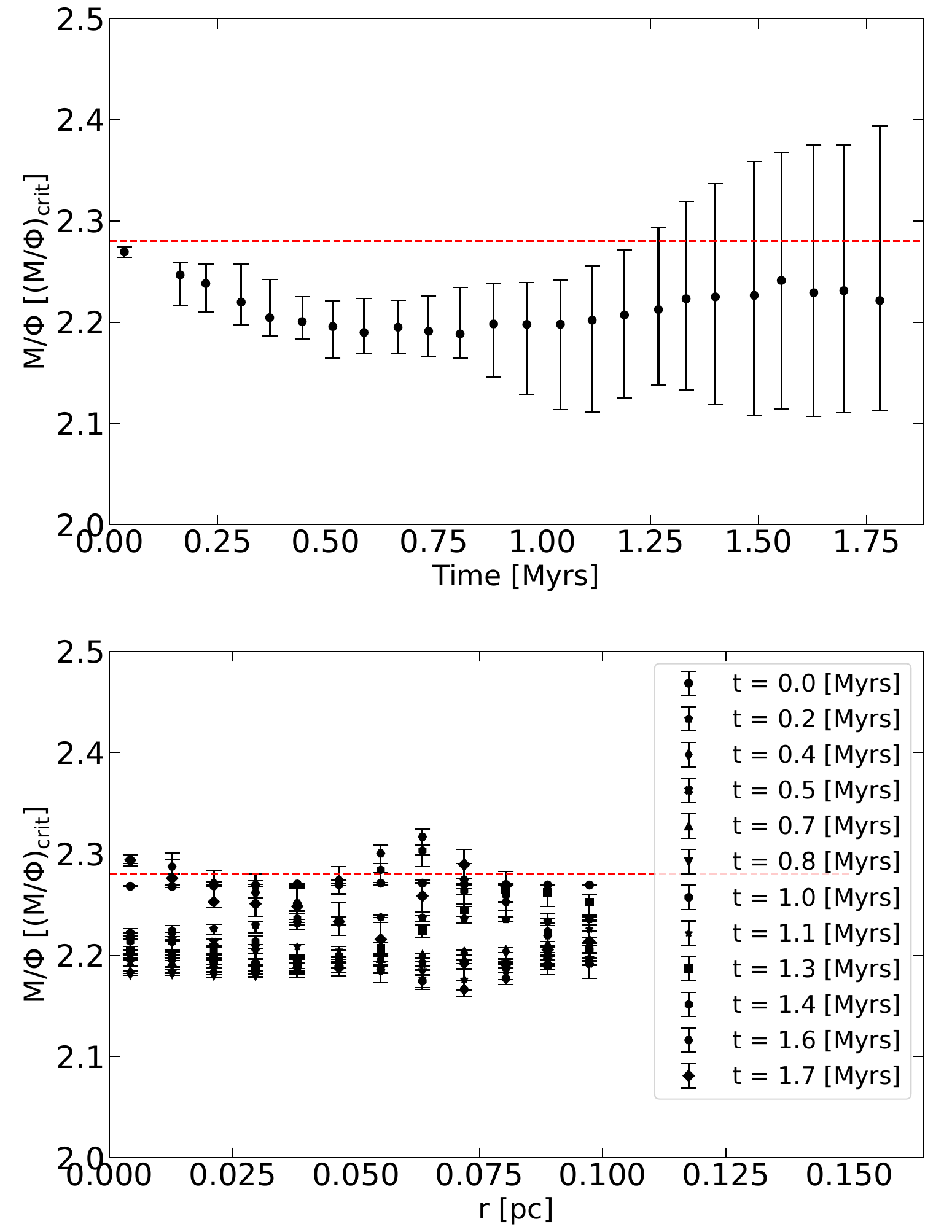}
\caption{Time evolution (upper panel) and radial profiles (lower panel) of the mass-to-flux ratio in an ideal MHD simulation. The red dashed line in both panels shows the initial value of the mass-to-flux ratio in the simulation. While there are some deviations from the original value (see Sect.~\ref{bench} for further explanation), the initial mass-to-flux ratio is recovered with notable accuracy. 
\label{MFBench}}
\end{figure}

We begin by defining a slice perpendicular to the $z$ axis that passes through the location of the maximum density at a time of $\sim$1.75 Myrs, when the central number density is $10^5~\rm{cm^{-3}}$. This slice spans 0.2 pc in both the $x$ and $y$ directions and its centered in the location of the maximum density. We compute the mass-to-flux ratio in this region, both at a time of $\sim$1.75 Myrs and also follow the evolution of the mass-to-flux ratio in the same region through time. The upper panel of Fig.~\ref{MFBench} shows this time evolution, where the points represent the 50th percentile from all pixels in the region, and the errorbars correspond to the 16th and 84th percentiles. The red dashed line in Fig.~\ref{MFBench} marks the initial condition of $\rm{M}/\Phi_B = $2.28. 

As is evident from the upper panel of Fig.~\ref{MFBench}, the measured mass-to-flux ratio is slightly below the initial value of $\rm{M}/\Phi_B = $2.28 for $t\lesssim$ 1 Myr (corresponding to 0.7 $t_{ff}$). This discrepancy is not necessarily a systematic bias as, due to our open boundary conditions and the initial turbulent conditions, some mass can leave the computational domain. At later times, however, the errorbars indicate that some cells exhibit larger mass-to-flux ratios than the initial condition. This is due to two factors. The first is some unavoidable numerical diffusion in the simulation itself. The second is that we are more prone to errors as the cloud is collapsing. At late times, as the cloud is collapsing along magnetic field lines, most of the mass is concentrated within a thin slice which is approximately perpendicular to the $z$ direction. As such, interpolation errors in only a handful of $dm_i$ elements are sufficient to overestimate or underestimate the mass-to-flux ratio. Despite these challenges, the measured values for the true mass-to-flux ratio remain very close to the initial value of 2.28 throughout the time evolution of the simulation.

In the lower panel of Fig.~\ref{MFBench} we present radially averaged profiles in concentric rings centered again around the location of the maximum density. The errorbars correspond to the 16th and 84th percentiles from all pixels in each concentric ring. Once more, the values measured are very close to the initial condition, albeit somewhat lower due to the same reasons explained above. Notably, there is very little variation as a function of radius, suggesting that no significant spatial biases are present in our mass-to-flux ratio measurements. Based on these tests, we conclude that the systematic errors on the measured mass-to-flux ratio for individual pixels, excluding changes due to nonideal MHD effects, are of the order of 0.2-0.3.

\subsection{Calculating the ``observed'' mass-to-flux ratio}\label{mockMF}

In contrast to the process described in Appendix~\ref{trueMF}, the ``observed'' magnetic flux, when the cloud is viewed along the $z$ axis, is calculated as
\begin{equation}
\Phi_{B,obs} = B_z(z_k)dx^2,
\end{equation}
where $dx$ is the resolution of each cell. The ``observed'' mass that is (erroneously) assigned to that ``flux tube'' is calculated as
\begin{equation}
\rm{M_{\Phi_B,obs}} = \sum_{z_k=0}^N \rho_{x_i, y_j, z_k}dx^3,
\end{equation}
where $N$ is the number of cells in the $z$ direction.

\section{Density and mass-to-flux ratio radial profiles at 1 $t_{ff}$}\label{RadProfs1tff}

Here, we present radial profiles of the cloud's number density, along with the true and observed mass-to-flux ratios, at a time of $1~t_{ff}$. However, a radial profile of the number density created in the same manner as for 1.44 $t_{ff}$ is ill defined for two reasons. Firstly, the ``denser'' structure (the maximum number density at this stage is $\sim7.5\times10^3~ \rm{cm^{-3}}$) is significantly tilted with respect to the principal axes of the grid (see the blue contours in the upper left and middle left panels of Fig.~\ref{Slices}). As such, considering a radial profile on the $x-y$ plane would misrepresent the cloud's number density. Secondly, the cloud has not yet fully thermally relaxed along magnetic field lines to form a flattened configuration with approximately uniform density along its minor axis. Therefore, a one-to-one correspondence between the radial profiles of the true mass-to-flux ratio and the number density should also not be expected since the former is more affected (compared to later times) from the gas distribution further away from the midplane. 

To provide a meaningful comparison with the results presented at later times, we constructed the radial profiles using a modified procedure. We first selected a subregion of the cloud, spanning 0.6 pc in each of the $x$, $y$, and $z$ directions (i.e., similar to the subregion used in Sect.~\ref{MFMain}), but now extended across all three dimensions. We then performed a weighted principal component analysis (PCA) to the density distribution in order to determine the orientation of the ``denser'' region within the cloud. In this context, by denser gas we define regions with density exceeding the 84th percentile of the density distribution within the selected subregion. As weights, we used the actual density values. We then rotated the full 3D density structure such that the minor axis of the denser region at this stage aligns with the $z$ direction. This allowed us to define a quasi-planar frame in which we averaged the density (within 0.1 pc) to mitigate density variations along the minor axis. To avoid contamination from interpolation artifacts near the edges of the rotated cube, we masked regions affected by zero-padding by assigning undefined values, which were then excluded from the radial averaging. The radial density profile was finally computed in concentric annuli centered on the peak of the averaged density. The same PCA-based rotation was applied to the 2D maps of the true and observed mass-to-flux ratios prior to creating the radial profiles so that there is one-to-one spatial correspondence with the density. 

\begin{figure}
\includegraphics[width=1.\columnwidth, clip]{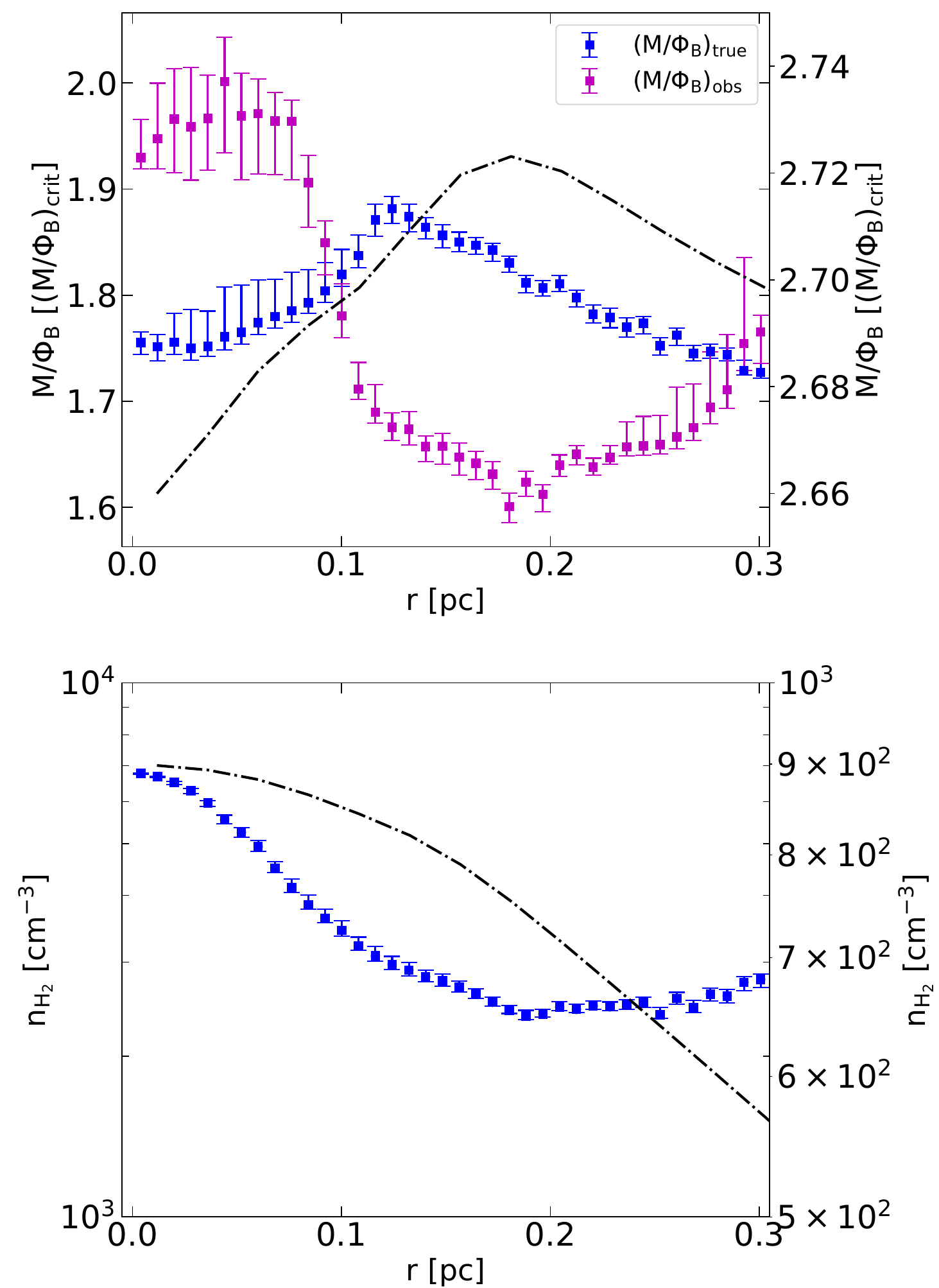}
\caption{Average profiles of the mass-to-flux ratio (upper panel), and the number density (lower panel) of the cloud in concentric radial shells at a time of 1 $t_{ff}$. With the blue points in the upper panel we show the true mass-to-flux ratio, and with the magenta points we show the observed mass-to-flux ratio. The black dash-dotted lines in the upper and lower panels show, respectively, radial profiles of the true mass-to-flux ratio and the density from a 2D axisymmetric supercritical simulation \citep{2023MNRAS.521.5087T}.
\label{AveragedMFearly}}
\end{figure}

We present our results on the radial profiles of the mass-to-flux ratio and the number density, in the upper and lower panels of Fig.~\ref{AveragedMFearly}, respectively. Radial profiles were calculated following the same procedure as in Sect.~\ref{MFMain}. That is, the points represent the 50th percentile from all pixels in each shell and the errorbars represent the 16th and 84th percentiles, respectively. Fig.~\ref{AveragedMFearly} reveals two interesting features. First, similarly to later times, the observed mass-to-flux ratio does not correspond exactly to the true one. This indicates that the projection effects in mass-to-flux ratio observational measurements, discussed in Sect.~\ref{MFMain}, are relevant throughout the cloud's evolution. Second, the radial profile of the true mass-to-flux ratio does not peak at the same location as the density. The difference in the value of the mass-to-flux ratio between the center and its peak is relatively small ($\sim$0.1–0.15), but nonetheless noticeable. As previously mentioned, this could be attributed to the fact that the true mass-to-flux ratio at this stage would be more affected by the density distribution further away from the midplane of the cloud. However, the fact that the true mass-to-flux ratio peaks $\sim$0.1 pc away from the peak location of the density could also be a consequence of the structure of the drift velocity at this, and previous, evolutionary stages. To explore this possibility, we compute the radial profile of the mass-to-flux ratio in a 2D axisymmetric supercritical simulation (i.e., model ``\texttt{M/$\Phi$2.6\_$\zeta/\zeta_01$\_$A_v$10\_T10}'' from \citealt{2023MNRAS.521.5087T}). The initial mass-to-flux ratio in this simulation (in units of the critical value) was 2.6 and the radial profile was calculated in the midplane of the cloud. The corresponding profile of the number density in the midplane of the cloud is shown with the black dash-dotted line in the lower panel of Fig.~\ref{AveragedMFearly}. Despite the significantly simpler morphology of this cloud, and the fact that only gravitationally-driven ambipolar diffusion was present, the mass-to-flux ratio is still not centrally peaked, although the effect is even smaller than the 3D simulation presented herein. At later times, the mass-to-flux ratio profile in the 2D simulation (not shown here) also peaks at the center of the cloud (i.e., at $r = 0$). These results suggest that the structure of the drift velocity in supercritical clouds could contribute to spatial offsets between the peak of the mass-to-flux ratio and that of the density.

\end{appendix}

\end{document}